\documentclass[twocolumn]{aastex631}

\usepackage[switch]{lineno}

\usepackage[utf8]{inputenc}
\usepackage{amssymb}
\usepackage{newunicodechar}

\usepackage{xcolor}
\newcommand{\jh}[1]{#1}

\begin{document}

\title{From inter-filamentary gas to filaments and hubs: gas flows in the Mon R2 hub–filament system}

\author[0000-0001-7866-2686]{Jihye Hwang}
\email{astrojhwang@gmail.com}
\affil{Institute for Advanced Study, Kyushu University, Japan}
\affil{Department of Earth and Planetary Sciences, Faculty of Science, Kyushu University, Nishi-ku, Fukuoka 819-0395, Japan}
\affil{Korea Astronomy and Space Science Institute (KASI), 776 Daedeokdae-ro, Yuseong-gu, Daejeon 34055, Republic of Korea}

\author[0000-0002-1959-7201]{Doris Arzoumanian}
\affil{Institute for Advanced Study, Kyushu University, Japan}
\affil{Department of Earth and Planetary Sciences, Faculty of Science, Kyushu University, Nishi-ku, Fukuoka 819-0395, Japan}

\author[0000-0001-9368-3143]{Yoshito Shimajiri}
\affil{Kyushu Kyoritsu University, Jiyugaoka 1-8, Yahatanishi-ku, Kitakyushu, Fukuoka, 807-08585, Japan}

\author[0000-0002-0963-0872]{Masahiro N. Machida}
\affil{Department of Earth and Planetary Sciences, Faculty of Science, Kyushu University, Nishi-ku, Fukuoka 819-0395, Japan}

\author[0000-0003-4366-6518]{Shu-ichiro Inutsuka}
\affil{Department of Physics, Graduate School of Science, Nagoya University, Furo-cho, Chikusa-ku, Nagoya, 464-8602, Japan}

\author[0000-0002-7066-4828]{M. S. N. Kumar}
\affil{Instituto de Astrof\'{i}sica e Ci\^{e}ncias do Espa\c{c}o, Universidade do Porto, CAUP, Rua das Estrelas, 4150-762 Porto, Portugal}

\author[0000-0003-4271-4901]{Shingo Nozaki}
\affil{Department of Earth and Planetary Sciences, Faculty of Science, Kyushu University, Nishi-ku, Fukuoka 819-0395, Japan}

\author[0000-0002-2062-1600]{Kazuki Tokuda}
\affil{Faculty of Education, Kagawa University, Saiwai-cho1-1, Takamatsu, Kagawa 760-8522, Japan}

\begin{abstract}
Hub-filament systems (HFSs) play an important role in the formation of massive stars and star clusters. Although the velocity structures along dense filaments have been studied, the gas kinematics in the low density inter-filament regions has not been investigated.
We use $^{13}$CO ($J$ = 1--0) and C$^{18}$O ($J$ = 1--0) observations obtained with the Nobeyama 45 m telescope to study the gas dynamics towards the Monoceros R2 (Mon R2) HFS. From the $^{13}$CO and C$^{18}$O data, tracing low- and high-density gas, respectively, we identify velocity coherent structures and divide them into filaments (Fs) and inter-filamentary regions (IFs). 
We estimate velocity gradients ($\Delta v$) and mass accretion rates ($\dot{M}$) along ($\parallel$) and across ($\perp$) the Fs and IFs. 
The mean ratio of $\dot{M}_\parallel$ to $\dot{M}_\perp$ in Fs is 6.8, while that in IFs is 1.5. These results show that the overall gas within both Fs and IFs flows directly into the hub and the gas flows faster along the Fs than the IFs. In addition, we found that at least 30\% of the gas mass in the IFs may flow towards the Fs replenishing the latter with new matter.
Our study reveals the importance of considering the total gas mass reservoir, both low- and high-density, infalling into the hub and promoting the formation of massive stars, which are preferentially located in the hub of Mon R2.

\end{abstract}

\keywords{Star Formation (1569) --- Interstellar Medium (847) --- Molecular clouds (1072) --- Massive stars (732)}


\section{introduction} \label{sec:intro}

Hub-filament Systems (HFSs) are ubiquitous in star-forming regions and play an important role in the formation and evolution of high-mass stars and star clusters \citep[e.g.,][]{Myers2009,Peretto2013,Fukui2019,Tokuda2019,Kumar2020,Tokuda2023}. In HFSs, filaments are elongated structures with high aspect ratios, whereas hubs, located at the junctions of filaments, show larger column densities and low aspect ratios (\citealt{Myers2009}; \citealt{Peretto2013}). In observations, longitudinal mass flows detected along filaments are suggested to enhance the density and mass of the hubs and consequently increase star-formation activity within them \citep{Peretto2013, Peretto2014, Chen2019, Trevino2019}. Indeed, high luminous sources ($>$ 10$^5$ M$_\odot$) in the galactic plane are preferentially found in HFSs \citep{Kumar2020}, indicating that HFSs provide a favorable environment for forming massive stars and star clusters. Although previous observational studies have demonstrated that the gas flowing along the filaments plays an important role in star formation in HFSs, it remains unclear whether the inter-filament gas, which is less dense than the gas along the filaments, also flows towards the hub, and how the mass accretion rates along the filaments compare with those in inter-filament regions.

To assess whether the mass accretion rate is high or low, information on the cloud mass and the gas dynamics is required. Monoceros R2 (Mon R2) is an excellent target for studying gas dynamics along the filaments as well as within the inter-filamentary environment, because it is one of the closest HFSs, at a distance of 830 pc \citep{Herbst1976} and  is at an early evolutionary stage ($\sim$1 Myr old).
Several filaments converge radially towards the central hub of Mon R2 \citep{Rayner2017, Trevino2019, Kumar2021}. 
The IRS 1 source associated with the ultracompact (UC) H{\sc ii} region is located at the center of the hub \citep{Herbst1976}, indicating the early stage of high-mass star-formation.
The physical, chemical, and magnetic properties of Mon R2 have been studied using the dust continuum \citep{Didelon2015, Rayner2017,Kumar2021}, molecular line \citep{Trevino2019}, and dust polarized emission  \citep{Hwang2022}. From the velocity gradients of C$^{18}$O (2--1), \citet{Trevino2019} inferred longitudinal gas flows along the filaments, transporting material toward the hub of Mon R2. They suggested a flattened structure of the Mon R2 hub with an inclination angle of 30$^\circ$ with respect to the plane of the sky. The magnetic field lines are aligned with filaments within the central 1 $\times$ 1 region of Mon R2 \citep{Hwang2022}, suggesting that magnetic fields may facilitate mass transport along the filaments toward the hub. Although mass flows along filaments have been investigated by previous studies \citep{Trevino2019}, the motion of the gas in the inter-filamentary regions has not been constrained.

To better understand the evolution of HFS and to estimate mass accretion flows from filaments and inter-filamentary regions onto hubs, we investigate the gas dynamics of the HFS Mon R2 using spectral line observations of C$^{18}$O ($J$ = 1 -- 0) and $^{13}$CO ($J$ = 1 -- 0) obtained with the Nobeyama 45 m telescope. These observations allow us to quantify gas flows in both dense and less dense regions tracing the filament and inter-filamentary regions, respectively. We present the observations and data reduction in Section \ref{sec:obs}. We show filaments and inter-filamentary regions obtained from velocity coherent structures extracted using a friends-of-friends algorithm applied to the multiple Gaussian fitting results of the C$^{18}$O and $^{13}$CO data in Section \ref{sec:res}. We also estimate the velocity gradients and mass accretion rates of each filament and inter-filamentary region. In Section \ref{sec:dis}, we discuss the mass accretion rate along and across the filaments and inter-filamentary regions of Mon R2 and the core mass function inside and outside the Mon R2 hub. Finally, we summarize our results and conclusions in Section \ref{sec:concl}.

\section{Observations} \label{sec:obs}

$^{13}$CO ($J$ = 1 -- 0) and C$^{18}$O ($J$ = 1 -- 0) spectral lines toward Mon R2 were observed with the Nobeyama 45 m telescope in 2021.
The On-The-Fly (OTF) maps were obtained using the FOREST receiver on the Nobeyama 45 m telescope.  
The C$^{18}$O and $^{13}$CO data cover a region of approximately 22$'$ $\times$ 22$'$ centered on IRS 1 in Mon R2 (5.3 pc $\times$ 5.3 pc at a distance of 830 pc). Their FWHM is 21.8$''$ ($\sim$0.09 pc) in both datasets. The rms noise level of both data is 0.3 K at a velocity resolution of 0.11 km s$^{-1}$.

\section{Analysis and Results}\label{sec:res}

\subsection{Filament and inter-filament regions}\label{sec:fila}

\begin{figure*}[htb!]
\epsscale{1.0}
\plotone{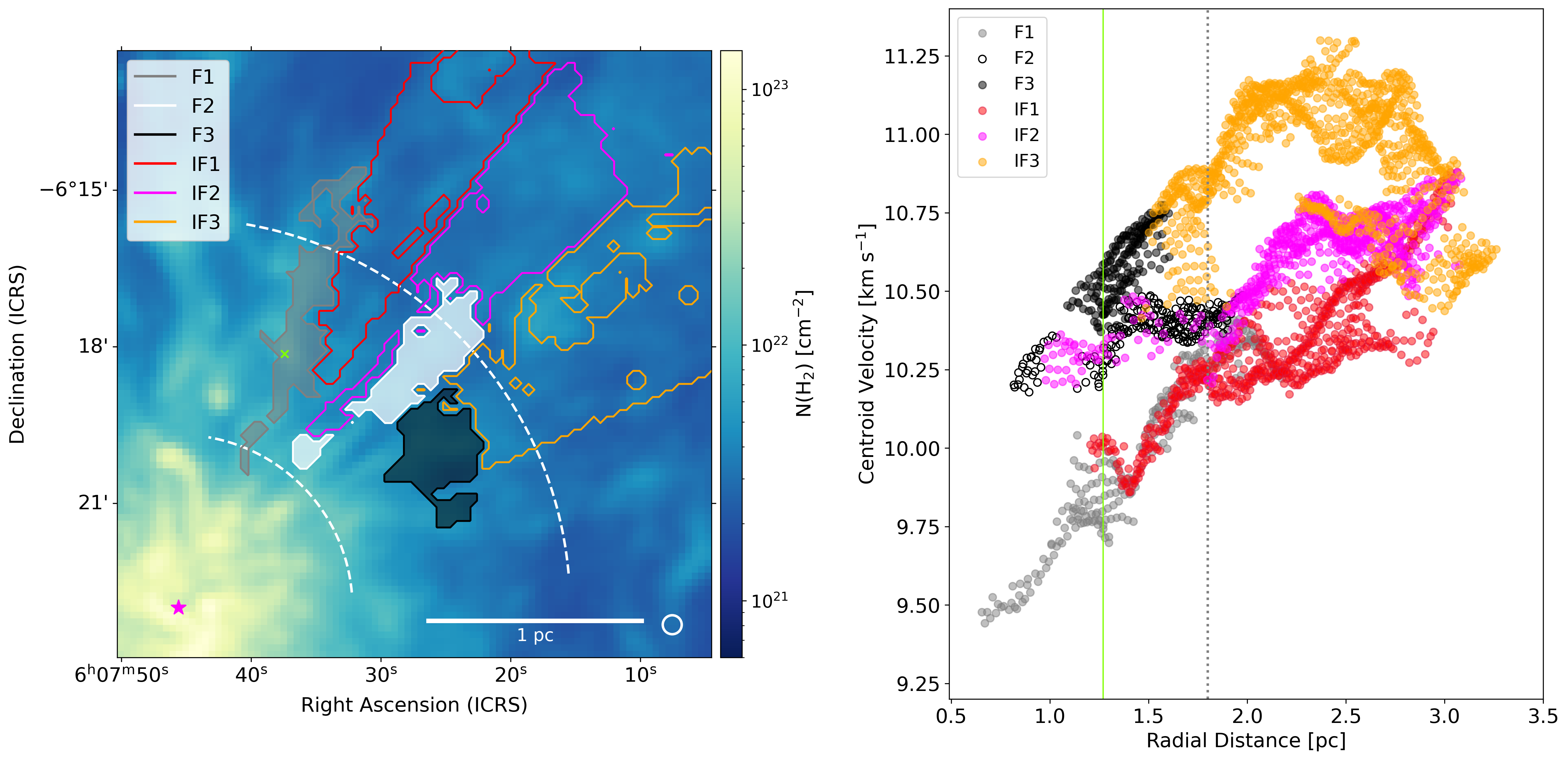}
\caption{(Left) Filaments (Fs) and inter-filamentary regions (IFs) overlaid on the H$_2$ column density map obtained from $Herschel$ data \citep{Didelon2015, Kumar2021}. The filled contours indicate individual regions identified as C$^{18}$O velocity coherent structures tracing the Fs. The non-filled contours correspond to the region identified as $^{13}$CO velocity coherent structures tracing the IFs. The two dotted lines indicate the radial distance of 0.8 pc and 1.8 pc from the magenta star indicating IRS 1, corresponding to the size of hub and the radial extent analyzed in this work, respectively. The scale bar and beam size of 21.8$''$ are shown at the bottom right corner. The green cross indicates the position of the prestellar core with the mass of 2.2 M$_\odot$ identified using $Hershecl$ data by \citet{Rayner2017}. (Right) Centroid velocities of $^{13}$CO components within each contour as a function of the radial distance from the central star shown in the left panel. The names of each contour region are labeled at the top left corner.
The green and gray lines show the position of the prestellar core and the 1.8 pc radial distance from IRS 1. 
\label{fig:velgrad}} 
\end{figure*}

Four {\sc{Hii}} regions are located in Mon R2, and their sizes and locations are described in \citet{Didelon2015}. The ionizing sources are marked as crosses in the figure of Appendix \ref{sec:temp}. The areas surrounding these sources show higher dust temperatures and lower column densities derived from $Herschel$ dust continuum data \citep{Didelon2015} observed as part of the Herschel imaging survey of OB Young Stellar objects (HOBYS) \citep{Motte2010}. 
In this paper, we focus on the velocity coherent structures in the north-west region of Mon R2, which is less affected by the {\sc{Hii}} regions and where we identified an ordered pattern of linear filament and inter-filament regions (Figure \ref{fig:velgrad}, see also Appendix \ref{sec:velstruc}). 

We identify velocity coherent structures from a multi-Gaussian fitting and friends-of-friends algorithm of the $^{13}$CO and C$^{18}$O position-position-velocity data cube. The details of identifying the velocity coherent structures are similar to the method used in \citet{Hwang2026}. We determine IRS 1 as the center of the Mon R2 hub and discuss the gas velocity structure towards the center of the hub. We define the filaments (Fs) and inter-filament (IF) regions using the $^{13}$CO and C$^{18}$O data. Because $^{13}$CO is about 5-33 times more abundant than C$^{18}$O \citep[e.g.,][]{Shimajiri2014}, it can trace more diffuse gas with low column densities. Consequently, the C$^{18}$O and $^{13}$CO data primarily probe dense filamentary structures and more diffuse inter-filamentary regions, respectively.
 The $^{13}$CO emission is optically thin in the region analyzed in this paper. 
In the north west region, we identified a single large velocity coherent structure in $^{13}$CO (cf. Appendix \ref{sec:velstruc}). Towards this $^{13}$CO extended velocity coherent structure, we find three spatially distinct and velocity coherent structures detected also in C$^{18}$O. We label these three C$^{18}$O detected structures as F1, F2, and F3. The remaining areas detected only in $^{13}$CO tracing the less dense inter-filament gas are labeled as IFs.

The right panel of Figure \ref{fig:velgrad} displays a position-velocity (PV) plot where the centroid velocities are shown as a function of the radial distance from the center of the hub. These three distinct structures in the PV diagram correspond to distinct regions on the two dimensional position-position map (see Figure \ref{fig:velgrad}-left).
We distinguish three regions with three ranges of position angles (IF1: 55$^\circ$-67$^\circ$, IF2: 42$^\circ$-55$^\circ$ IF3: 23$^\circ$-42$^\circ$). When dividing the three IFs based on this criterion, a few pixels that are spatially connected to IF1 are classified as belonging to IF2. Although we adjusted  the ranges of position angles, some pixels still remained ambiguously assigned to IF1 or IF2. To address this, we used an additional criterion, where pixels in IF2 are required to have centroid velocities greater than 10.2 km s$^{-1}$, the smallest velocity of the bulk of IF2.
The 3 Fs and 3 IFs identified following the above procedure are indicated in the left panel of Figure \ref{fig:velgrad} as filled and unfilled contours, respectively. 
Inside the hub within the 0.8 pc, velocity coherent structures have sizes smaller than three beams (see Appendix \ref{sec:velstruc}). In this paper, we focus the analysis on the gas kinematics outside the hub.
In the right panel of Figure \ref{fig:velgrad}, the slopes of the centroid velocities change at a distance of 1.8 pc, especially in IF1 and IF2. Moreover, velocities outside 1.8 pc along IF1 and IF2 are more complex. We cannot exclude the possibility of the velocity connection between IF1 and IF2 (Figure \ref{fig:velgrad}). 
Additionally, The Fs detected from C$^{18}$O extend to a $\sim$1.8 pc radial distance from the hub. Therefore, in this paper, we focus on the analysis of Fs and IFs in the radial distance between 0.8 pc and 1.8 pc. The analysis for the whole region will be investigated in a further paper.
The physical parameters derived within this range of radial distance, including velocity gradients and mass flow rates for Fs and IFs, are summarized in Table \ref{tab:para}.

\begin{deluxetable*}{ccccccccccc}
\tablecaption{Physical parameters of the filaments (F) and inter-filament (IF) regions \label{tab:para}}
\tablecolumns{11}
\tablenum{1}
\tablewidth{0pt}
\tablehead{(1) & (2) & (3) & (4) & (5) & (6) & (7) &(8) & (9) & (10) &(11)\\
\colhead{Name} &
\colhead{$M_{^{13}\mathrm{CO}}$} &
\colhead{$M_{\mathrm{H}_2}$} & \colhead{$\Delta v_{\parallel \mathrm{,all}}$} & \colhead{$\Delta \overline{v_\perp}$} & \colhead{$L$} & \colhead{$\dot{M}_{\parallel \mathrm{,all}}$} & \colhead{$\dot{M}_{\perp \mathrm{,all}}$} & \colhead{$\dot{M}_{\parallel}$} & \colhead{$\dot{M}_\perp$} & \colhead{$\overline{R}$}
\\ &  [$10^{-4}\, M_\odot$] & [$M_\odot$] & \multicolumn{2}{c}{[km s$^{-1}$ pc$^{-1}$]} & [pc] & \multicolumn{2}{c}{[$M_\odot$ Myr$^{-1}$]}& \multicolumn{2}{c}{[$M_\odot$ Myr$^{-1}$]}}
\startdata
F1 & \jh{4.7} & \jh{26.8} & 0.76 $\pm$ 0.02 & 0.34 (0.16) & 1.5 & \jh{20.8} & \jh{9.7} & \jh{2.1}--\jh{9.8} (\jh{5.4}) & \jh{0.3}--\jh{3.8} (\jh{1.9}) & 7.7 (3.9) \\
F2 & \jh{4.0} & \jh{22.7} & 0.20 $\pm$ 0.01 & 0.29 (0.24) & 1.2 & \jh{17.7} & \jh{6.5} & \jh{0.8}--\jh{6.3} (\jh{2.5}) & \jh{0.1}--\jh{4.8} (\jh{1.3}) & 5.3 (2.9) \\
F3 & \jh{4.4} & \jh{25.2} & 0.70 $\pm$ 0.04 & 0.21 (0.16)  & 0.5 & \jh{20.0} & \jh{3.5} & \jh{0.8}--\jh{14.2} (\jh{7.9}) & \jh{0.8}--\jh{1.7} (\jh{1.2}) & 7.5 (3.7)\\
IF1 & \jh{2.7} & \jh{15.3} & 0.64 $\pm$ 0.03 & 0.46 (0.57) & 1.9 & \jh{11.9} & \jh{10.6} & \jh{2.4}--\jh{8.8} (\jh{4.7}) & \jh{0.6}--\jh{4.9} (\jh{2.7}) & 2.3 (0.7) \\
IF2 & \jh{1.5} & \jh{8.5} & 0.16 $\pm$ 0.03 & 0.47 (0.21) & 1.0 & \jh{6.6} & \jh{4.4} & \jh{0.2}--\jh{1.7} (\jh{1.0}) & \jh{0.4}--\jh{1.9} (\jh{1.1})  & 1.2 (0.5) \\
IF3 & \jh{2.7} & \jh{15.2} & 0.26 $\pm$ 0.13 & 0.62 (0.07) & 1.2 & \jh{11.8} & \jh{13.3} & \jh{0.4}--\jh{4.5} (\jh{2.4}) & \jh{2.0}--\jh{11.3} (\jh{6.7}) & 1.1 (1.1) \\
\enddata
\tablecomments{Columns: (1) Name of velocity coherent structure (Figure \ref{fig:velgrad}) (2) $^{13}$CO mass. (3) H$_2$ mass \jh{(see detail calculations in Appendix \ref{sec:temp})}. (4) Velocity gradient along the structures derived from the linear fits of the velocities from 0.8 pc to 1.8 pc and the uncertainties of the fit. (5) Mean absolute velocity gradient across the sub-structures (along the arcs, see Appendix \ref{sec:velstruc}). The value in parentheses indicate the standard deviation of the values. (6) Length of the structure along the radial direction from the minimum radial distance of each F or IF to 1.8 pc. (7) Mass accretion rate along the structure calculated using the columns (3) and (4). 
(8) The sum of mass accretion rate across the sub-structures of the column (10).
(9) Range of mass accretion rates along the sub-structure, measured in steps of 0.2 pc in radial distance (Appendix \ref{sec:arc}). The value in parentheses indicates the mean of the range. (10) Range of mass accretion rates across the sub-structure, measured in steps of 0.2 pc in radial distance. The value in parentheses indicates the mean of the range. (11) Mean ratio of the mass accretion rates along and across the structure ($\dot{M}_{\parallel}$/$\dot{M}_{\perp}$) using columns (9) and (10), measured in steps of 0.2 pc in radial distance. The value in parentheses shows the standard deviation of the values.}
\end{deluxetable*}

\subsection{Velocity Gradients}\label{sec:velgra}

We estimate velocity gradients of Fs and IFs both toward the hub ($\parallel$, radial direction) and across the structures ($\perp$, azimuthal direction in the plane of the sky). 
The Fs, detected in both $^{13}$CO and C$^{18}$O, show consistent velocities in both isotopologues. We therefore use the $^{13}$CO data to calculate the velocity gradients in both Fs and IFs. 

The velocity gradients are estimated by linear fits to the $^{13}$CO centroid velocities as a function of the radial distance between 0.8--1.8 pc. The velocity gradients along the Fs and IFs ($\Delta v_{\parallel ,\mathrm{all}}$) derived from the fits are within the range of 0.20--0.76 km s$^{-1}$ pc$^{-1}$ and 0.16--0.64 km s$^{-1}$ pc$^{-1}$, respectively (Table \ref{tab:para}). The sign of $\Delta v_{\parallel,\mathrm{all}}$ is always positive. The velocity gradients across the Fs and IFs ($\Delta v_{\perp}$) are measured by dividing them into sub-regions with a radial spacing of 0.2 pc (Appendix \ref{sec:arc}). If the number of pixels included in a given sub-region is larger than one beam size (6 pixels), we linearly fit the centroid velocities as a function of position angle along the arc (see details in Appendix \ref{sec:arc}). The mean velocity gradients across the entire extent of the Fs and IFs ($\Delta \overline{v_\perp}$) have ranges of 0.21--0.34 km s$^{-1}$ pc$^{-1}$ and 0.46--0.62 km s$^{-1}$ pc$^{-1}$, respectively. These ranges are absolute values. The general trends and the comparison of the local velocity gradients along and across Fs and IFs are discussed in Section \ref{sec:acc}.

\subsection{Mass Accretion Rates}\label{sec:massacc}

We calculate the mass accretion rates along ($\dot{M}_{\parallel}$) and across ($\dot{M}_{\perp}$) the Fs and IFs. The mass accretion rate ($\dot{M}$) is calculated as
\begin{equation}
\dot{M}=M_{\mathrm{H}_2}\, \Delta v \, /\tan{i}, 
\end{equation}
where $M_{\mathrm{H}_2}$ is the total mass of the molecular hydrogen gas estimated from the $^{13}$CO column densities (Appendix \ref{sec:temp}), $\Delta v$ is the absolute velocity gradient, and $i$ is the inclination angle with respect to the plane of the sky with 0$< i <$90 degree. We assume $i$ = 45$^\circ$. If the inclination angle is smaller than 45$^\circ$, the mass accretion rate will increase with the factor of 1/$\tan{i}$. The $^{13}$CO column densities, the total gas masses, the velocity gradients, and the mass accretion rates for Fs and IFs are listed in Table \ref{tab:para}. 

The $\dot{M}_{\parallel, \mathrm{all}}$ within the radial distance between 0.8 pc and 1.8 pc are \jh{17.7--20.8} M$_\odot$ Myr$^{-1}$ and \jh{6.6--11.9} M$_\odot$ Myr$^{-1}$ in Fs and IFs, respectively. We estimate the local $\dot{M}_{\perp}$ in sub-regions with a radial spacing of 0.2 pc (as explained in Section \ref{sec:velgra} and Appendix \ref{sec:arc}). The $\dot{M}_{\perp, \mathrm{all}}$ is estimated by integrating the values of $\dot{M}_\perp$, which are \jh{3.5--9.7} M$_\odot$ Myr$^{-1}$ and \jh{4.4--13.3} M$_\odot$ Myr$^{-1}$ in the Fs and IFs, respectively. We note that the total masses used to calculate $\dot{M}_{\parallel, \mathrm{all}}$ and $\dot{M}_{\perp, \mathrm{all}}$ are different since the sub-regions having the number of pixels less than one beam size (6 pixels) are excluded in the calculation of $\dot{M}_{\perp}$.  
In addition, we note that all $\dot{M}$ values are estimated from the absolute velocity gradients and therefore represent the amount of gas flow, not its direction. Interestingly, the overall sign of $\Delta v _\perp$ along IF1 and IF2 is opposite, suggesting opposite direction of the flow.   
Although, because of projection effect and the line of sight measurement of the velocity, we cannot specify the direction of the gas flows, a reasonable assumption is  that the flow is gravitationally directed towards  the nearby dense filament, with the gas in IF1 and IF2 directed towards F1 and F2, respectively.

To compare the mass accretion rates along and across each F or IF, we also estimate the local $\dot{M}_{\parallel}$ in the same sub-regions, used to measure the local $\dot{M}_{\perp}$. The ranges of local mass accretion rates are listed in the columns (7) and (8) in Table \ref{tab:para}. 
In addition, we derive the local ratio of $\dot{M}_{\parallel}$ to $\dot{M}_{\perp}$ in each sub-region. The local ratio is independent of the mass and depends only on the ratio of local velocity gradients along and across Fs or IFs. We compare and discuss the parameters derived for the Fs and IFs in Section \ref{sec:acc}.

\section{Discussions}\label{sec:dis}

\subsection{Mass accretion rate along and across filaments and inter-filament regions}\label{sec:acc}

We estimate the total mass accretion rate along Fs and IFs within the radial distance from 0.8 pc to 1.8 pc. 
The total mass accretion rate along Fs and IFs varies from \jh{6.6} to \jh{20.8} M$_\odot$ Myr$^{-1}$ (Column (7) in Table \ref{tab:para}). \citet{Trevino2019} found filamentary structures in Mon R2 using $^{13}$CO data obtained with the Institut de Radioastronomie Millim\'{e}trique (IRAM) 30 m telescope.
They identified nine main filaments on their entire map. The `F1' and `F9' filaments in their work are located close to the region we analyzed in this work. For these two filaments, they estimated mass accretion rates of 73 and 60 M$_\odot$ Myr$^{-1}$, respectively. Their mass accretion rates are \jh{3.5 and 3 times higher than our values found along the F1 and F3 ($\dot{M}_{\parallel ,\mathrm{all}}$). We cannot directly compare our values to theirs, because the lengths of our F1 and F3 are five times shorter than those of `F1' and `F9'. Additionally, the hub regions are included in the calculation of \citet{Trevino2019}. Due to these differences, our estimated mass accretion rates are smaller than their values. }

In addition to measuring the mass accretion rates along Fs as previously done \citep[e.g.,][]{Trevino2019}, we also derive those in IFs. 
The mass accretion rates for the full Fs and IFs are obtained towards different areas and thus different masses (See Table \ref{tab:para}). To compare the mass accretion rates for the same amount of gas, we estimate the local values and compute the ratios $\overline{R}=\dot{M}_\parallel/\dot{M}_\perp$. 
By deriving the ratios, we can directly compare the relative importance of mass accretion rates in the radial and azimuthal directions among sub-regions.

\begin{figure*}[htb!]
\epsscale{1}
\plotone{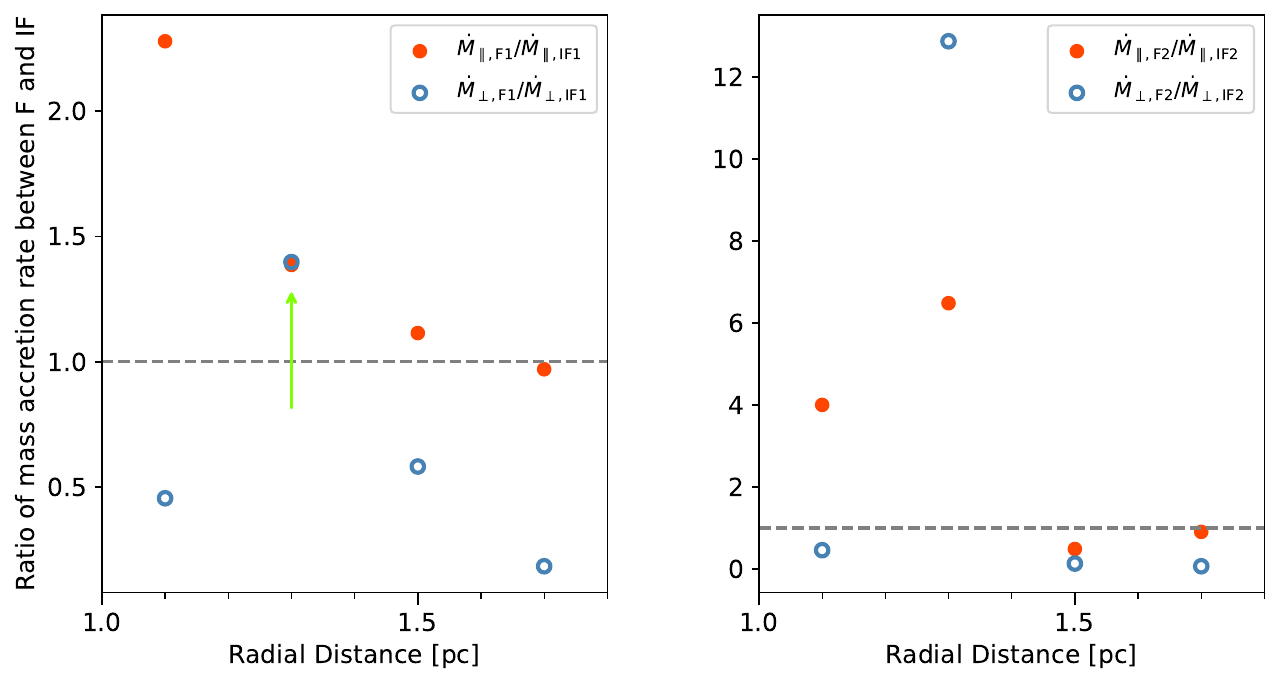}
\caption{Ratio of local mass accretion rate between F and IF along ($\parallel$, red circles) and across ($\perp$, blue circles) the structure. The left and right panels show the ratios between F1 and IF1, and F2 and IF2, respectively. The green arrow indicates the sub-region containing the prestellar core shown in Figure \ref{fig:velgrad}. The dashed lines indicate a ratio of one. \label{fig:FIF}} 
\end{figure*}

We compare the ratio of $\dot{M}_\parallel$ to $\dot{M}_\perp$ between F1 and IF1, and between F2 and IF2 as a function of the radial distance from the center of the hub in Figure \ref{fig:FIF}. We find similar trends in both cases. Most of the $\dot{M}_\parallel$ values in F1 and F2 are larger than those in IF1 and IF2, while most of the $\dot{M}_\perp$ values in F1 and F2 are smaller than those in IF1 and IF2. 
 The exception to this observed trend is the ratio derived in the sub-region containing the prestellar core, where $\dot{M}_{\perp,F1}$ is larger than $\dot{M}_{\perp,IF1}$. 
 The ratios of $\dot{M}_\parallel$ between Fs and IFs increase closer to the hub, while those of $\perp$ between Fs and IFs are mostly uniform along the radial distance. The mass accretion rates along F1 and F2 become more dominant compared to those of IF1 and IF2 as the radial distance from the hub decreases.
The ratio of $M_{\mathrm{H}_2}$ between F1 and IF1 increases with decreasing radial distance from the hub, from \jh{0.5} to \jh{4.7}, whereas that between F2 and IF2 decreases from \jh{7.8} to \jh{0.2}. The ratio of $\Delta v_\parallel$ between F1 and IF1 remains comparable in the range of radial distances, while that between F2 and IF2 increases with decreasing radial distance from the hub.
The combined variations of $M_{\mathrm{H}2}$ and $\Delta v_\parallel$ lead to an increase in the ratio of $\dot{M}_\parallel$ between Fs and IFs with the increase of the ratio of $\dot{M}_\parallel$ and $\Delta v_\parallel$ for F1-IF1 and F2-IF2, respectively (see also the results shown in Appendix \ref{sec:arc}). If the flow of matter is accelerated towards the hub by the gravity of the hub, we would expect an increase in the velocity gradient when approaching the hub. We can see such an increase in IF1 and F2, but we do not see it for F1 and IF2 (Appendix \ref{sec:arc}). The different behavior of the velocity gradient variations towards the hub in the different Fs and IFs may indicate distinct behavior of the matter flow and/or be affected by projection effects.
 
As shown in Figure \ref{fig:velgrad}, F3 and IF3 appear to be connected spatially and kinematically along the radial direction. The H$_2$ and $^{13}$CO column densities also gradually increase with decreasing distance from the hub. This may suggest that gas flows from IF3 to F3 building up more matter within F3, leading to a relatively higher density in F3, which is consequently detected in C$^{18}$O, unlike IF3 only detected in $^{13}$CO. We suggest that F3 and IF3 may belong to a continuous elongated structure connected to the hub, and thus the definition of F3 may differ from that of F1 and F2. Consequently, we do not compare the ratio of the mass accretion rates of IF3 and F3 since they do not show overlapping regions in the azimuthal direction (Appendix~\ref{sec:arc}).

\begin{figure*}[htb!]
\epsscale{0.8}
\plotone{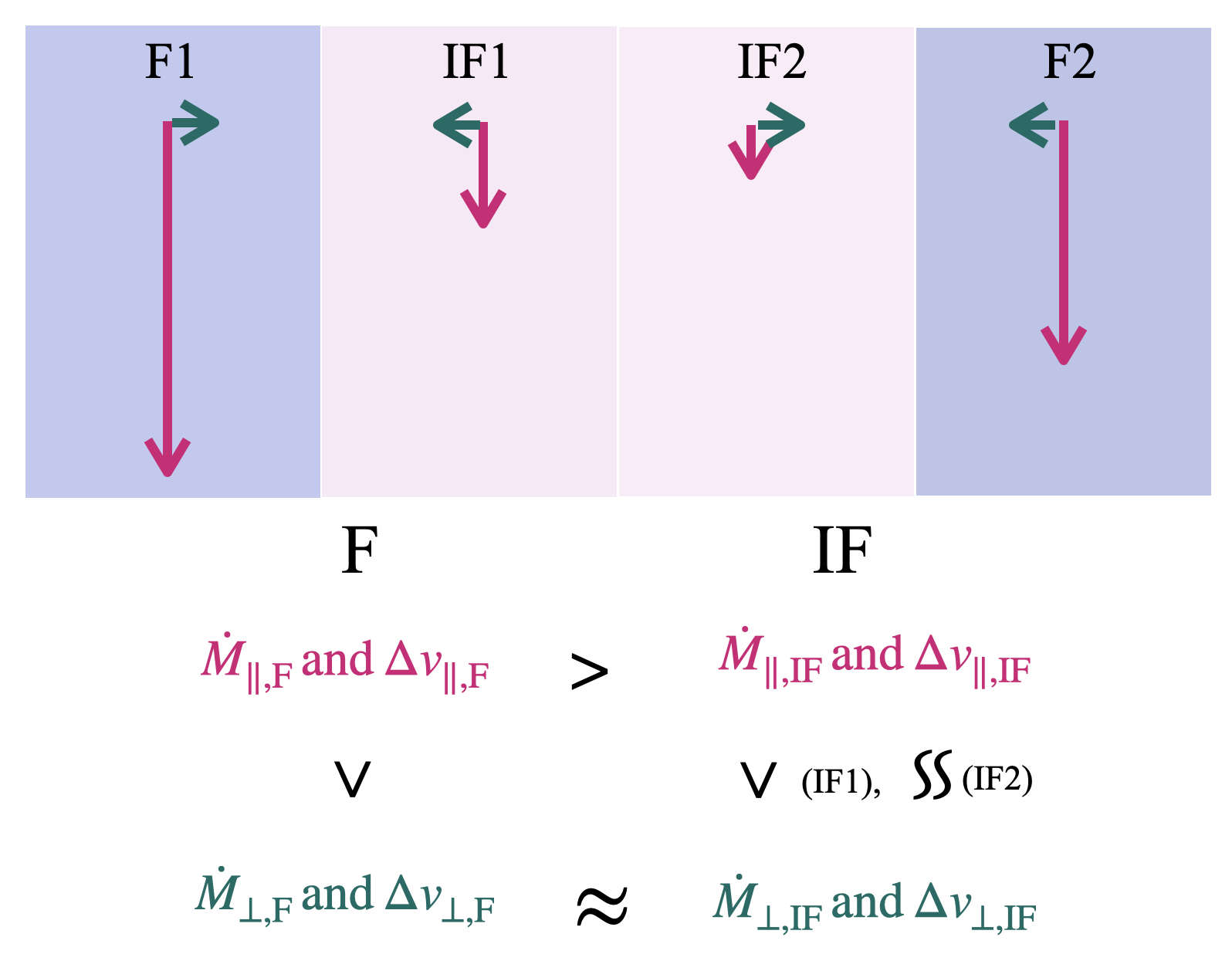}
\caption{Summary results of relative mass accretion ratio and velocity gradient along and across the F and IF. Top four boxes indicate each structure. The green and magenta arrows indicate the relative velocity gradient along and across the structure. The lengths of the arrows indicates the relative scale based on the mean ratio listed in Table \ref{tab:para}. The bottom chart summarizes the inequality of the mass accretion rates and velocity gradients across and along the F and IF based on the values in columns (9) and (10) of Table \ref{tab:para}. \label{fig:sum}} 
\end{figure*}

Figure \ref{fig:sum} summarizes the results of the mean local velocity gradients ($\Delta v$) and mass accretion rates ($\dot{M}$) along ($\parallel$) and across ($\perp$) the pairs F1, IF1 and F2, IF2 (columns (9), (10), and (11) in Table \ref{tab:para}). 
The relative mean values of $\Delta v$ and $\dot{M}$ show almost the same trend between the later Fs and IFs.
Except for IF2, $\dot{M}_\parallel$ ($\Delta v_\parallel$) is greater than $\dot{M}_\perp$ ($\Delta v_\perp$) in both Fs and IFs. In IF2, $\dot{M}_\perp$ ($\Delta v_\perp$) is comparable to $\dot{M}_\parallel$ ($\Delta v_\parallel$). 
$\dot{M}_\parallel$ in Fs are larger than those in IFs, while $\dot{M}_\perp$ in Fs are comparable to those in IFs.
A recent study using MHD simulations discussed the flow rate along and across filamentary structures and found that the mean flow rate along the structures is three times higher than across the structures \citep{Wells2025}. Our observational results are also compatible with those derived from HFS evolution simulations of \citet{Suin2025}, where filaments exhibit predominantly faster flows than their surrounding gas, and filament flows become faster as they approach the central hub. 

We further analyze the mass accretion rate in IF1 and F1 to discuss the gas flow near the prestellar core with a mass of 2.2 M$_\odot$ identified in F1 using the $Herschel$ data \citep{Rayner2017}. 
Both local $\dot{M}_\parallel$ and $\dot{M}_\perp$ at radial distances of 1.2--1.4 pc near the prestellar core in IF1 are comparable, with values of $\sim$\jh{3.8--4.0} M$_\odot$ Myr$^{-1}$ (Appendix~\ref{sec:arc}). 
However, in other sub-regions of IF1, the mean $\dot{M}_\parallel$ is 2.8 times higher than the mean $\dot{M}_\perp$.
This trend is also found in F1, the local $\dot{M}_\parallel$ and $\dot{M}_\perp$ are comparable, with values of $\sim$\jh{2.8--2.9} M$_\odot$ Myr$^{-1}$ in the sub-region including the prestellar core. However, the mean $\dot{M}_\parallel$ in other sub-regions is 9.8 times higher than $\dot{M}_\perp$. \jh{These results suggest that the prestellar core may play a dynamically important role in the immediate surrounding gas due to its self-gravity, while the large mass of the hub ($>$1500 M$\odot$; \citealt{Kumar2021}) may drive the overall gas flows toward the hub.} The mass infall rate of the prestellar core estimated from the free-fall time is 68 M$_\odot$ Myr$^{-1}$ (Appendix \ref{sec:infall}). However, the observed lifetime is usually 2-5 times longer than the free-fall time due to additional supports by turbulence, magnetic field, and thermal pressure \citep[e.g.,][]{Kirk2005, Andre2009}. \jh{Even when considering longer lifetime, the estimated mass infall rate of the prestellar core of 12.6--34 M$_\odot$ Myr$^{-1}$ remains higher than the local mass accretion rates measured in F1 and IF1 (2.8--2.9 and 2.0--2.1 M$_\odot$ Myr$^{-1}$, respectively). This comparison suggests that most of the gas  infalling onto the future protostar may originate from the prestellar core, with a small fraction infalling/accreted from the local material in the parent filament and inter-filament region. Prestellar cores may thus locally affect the gas motions and redirect part of the gas flow, while the bulk of the matter flows onto the hub.}


\subsection{Core Mass Function}

\begin{figure*}[htb!]
\epsscale{1.1}
\plotone{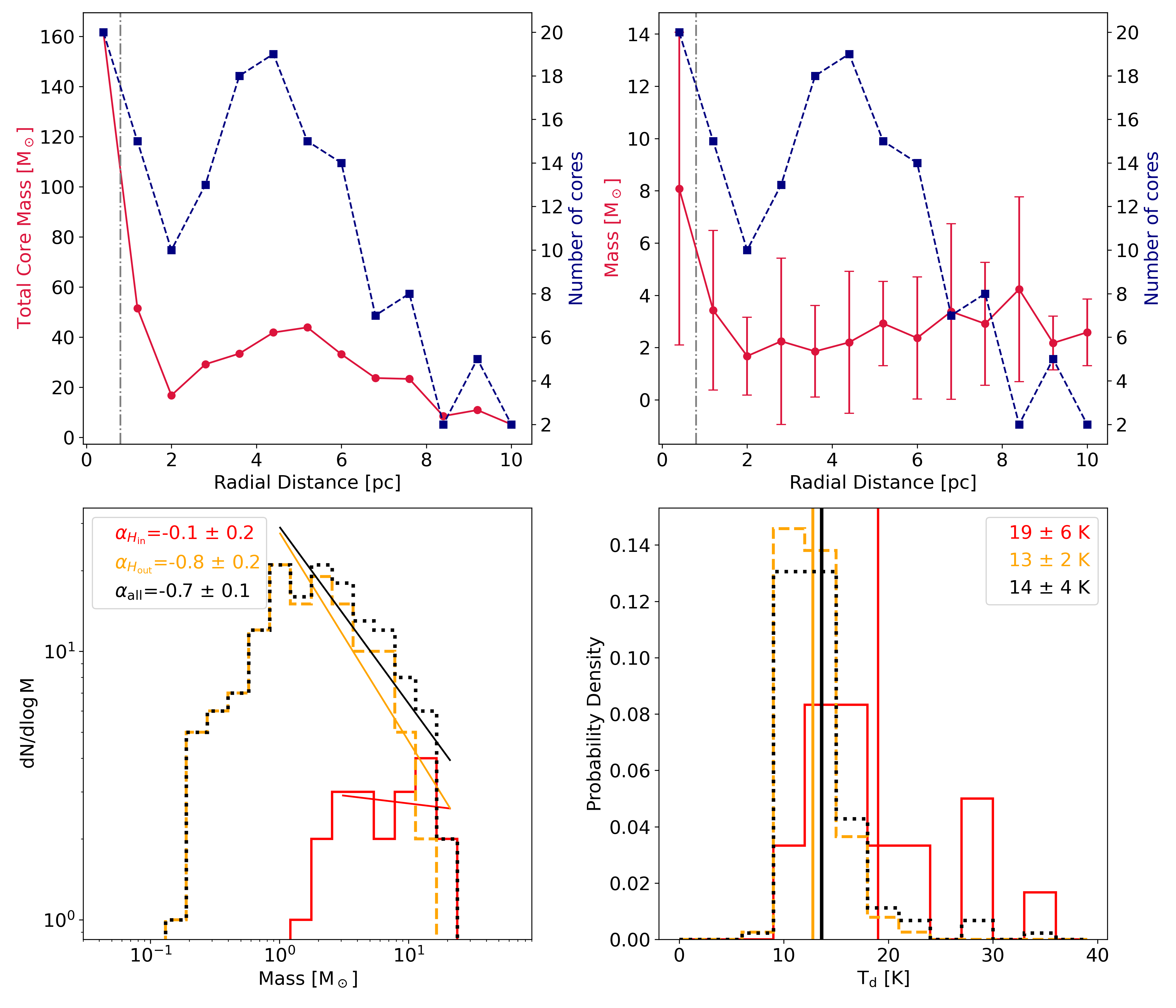}
\caption{The total (red line in the upper left panel) and the mean (red line in the upper right panel) core mass as a function of the radial distance from the IRS 1 in steps of 0.8 pc. The blue lines in both panels indicate the number of cores as a function of the radial distance. The error bars in the right panel indicate the standard deviation of the core mass within the 0.8 pc bin used to derive the mean core mass. The gray vertical lines mark a radial distance of 0.8 pc of the hub size in both panels. (Lower Left) Core mass function (CMF) in Mon R2 from the $Herschel$ catalog of \citet{Rayner2017}. The black dashed line shows the CMF for the 148 prestellar cores. The red and orange lines indicate the CMF within and outside the hub of Mon R2 (20 cores in the hub and 128 cores outside the hub). We used the same boundary for the hub region as \citet{Kumar2021}. They determined the hub region as the region interior to 0.8 pc from the central star shown in Figure \ref{fig:comp}. The power law fitting parameters are written at the right upper corner. (Lower Right) Distribution of dust temperature of cores inside and outside the hub (red and orange lines). The black dotted line indicates distribution of the dust temperature for all the cores. Vertical lines correspond to the mean values, which are given in the upper right of the panel. \label{fig:coremass}} 
\end{figure*}

To further study the impact of matter flow towards the hub, we investigate the distribution of core masses within and outside the hub, where we identify the hub as a circle centered on the IRS 1  with a radius of 0.8 pc \citep{Kumar2021}. For this purpose, we use the source catalog derived from $Herschel$ observations by \citet{Rayner2017}, which  identified a total of 176 cores. Among them, 28 cores are protostars mostly located in the hub. We derive the core mass function (CMF) of the prestellar cores inside and outside the hub. Out of a total of 148 cores, 20 (14\%) are located inside and 128 (86\%) are outside (Figure \ref{fig:coremass}). The largest number of cores in the figure is found inside the hub, where 20 cores are identified.
 The upper panels of Figure \ref{fig:coremass} show the total and mean core masses as a function of the radial distance from the IRS 1.
The mean core mass within the hub is approximately three times greater than that outside the hub. 
 
The lower left panel of Figure \ref{fig:coremass} shows the CMFs inside and outside the hub. The slopes of the high-mass end of the CMF differ significantly. The slope within the hub of 0.1 $\pm$ 0.1 (derived by fitting the cores with masses between 3 and 20 M$_\odot$) is shallower than that outside the hub of 0.8 $\pm$ 0.1 (derived by fitting the cores with masses between 1 and 20 M$_\odot$), indicating an excess of high-mass cores in the hub. Including protostars in the CMF makes the slope within the hub three times steeper (0.3 $\pm$ 0.1), while the CMF outside the hub remains unchanged.
 
 The lower right panel of Figure \ref{fig:coremass} shows the distribution of dust temperature of the cores. The mean dust temperatures of the the cores inside the hub are $\sim$1.5 times higher than those outside the hub. 
 \citet{Seshadri2024} have also reported that the cores in the hub are more massive and warmer than those outside the hub in the HFS of RCW 117. They suggested that the cores in the hub are in a more evolved stage. 
 We found that the cores inside the Mon R2 hub are on average more massive and warmer than them outside the hub. This oberved segregation of the core temperature suggests that the present star formation in the hub is affected by the stellar feedback, where the gas in the hub is heated by the more evolved stars (i.e., IRS 1, 2, 3, 4, and 5), previously formed in the hub. The observed segregation of the core mass suggests that the Mon R2 hub provides the conditions necessary to form higher mass cores compared to the surrounding filaments, as already described in previous works \citep[e.g.,][]{Peretto2013,Peretto2014,Kumar2020,Kumar2021}.

\subsection{Role of low and high density gas flows in the evolution of hubs}

We estimate mass accretion rates in the north-west part of the HFS, corresponding to a cone with an angle of $\sim$59 degree. The total mass accretion rate in the studied region is estimated to be \jh{88.8 M$_\odot$ Myr$^{-1}$} by integrating $\dot{M}_{\parallel,\mathrm{all}}$ in Fs and IFs. Assuming that the total mass accretion rate in this region is representative of the HFS, the total longitudinal inflow into the hub along Fs and IFs for the full Mon R2 region (scaled over 360 degree) is on the order of $\sim$\jh{542} M$_\odot$ Myr$^{-1}$. \jh{If we only consider the dense gas flow in Fs, the total mass accretion rate will be $\sim$1.5 times smaller.} 



We also find that a significant portion of  \jh{at least 30\%} of the inter-filamentary gas moves across the inter-filamentary regions to the filament, especially in IF2. 
Some studies discussed gas flow from (lower density) sub-filaments to (higher density) main filaments. In Mon R2, \citet{Trevino2019} identified sub-filaments connected to the main-filaments and measure one order of magnitude lower mass accretion rate along the sub-filaments compared to the main filaments. Simulations have discussed that most sub-filaments are in subcritical states, with mass flows on the order of tens of M$_\odot$ Myr$^{-1}$ \citep[e.g.,][]{Smith2016}. 

Previous studies have discussed the gas flow along dense structures \citep[e.g.,][]{Peretto2013,Liu2016,Hu2021,Yang2023,Zhang2024}. Our study suggests that considering only the mass accretion rate along the dense filaments will underestimate the total gas mass reservoir that could contribute to increasing the total mass of the hub. Our data further reveal that the diffuse gas does not flow solely towards the hub, but some of it first converges towards the filaments, becomes dense, and then flows along the filaments onto the hub.

In this paper, we focus on the northeastern region of the Mon R2 HFS. A comprehensive analysis of all velocity-coherent structures, including the kinematics within the hub (Appendix \ref{sec:velstruc}) will be presented in a future paper.

\section{Summary}\label{sec:concl}
 
We analyzed the C$^{18}$O ($J$ = 1--0) and $^{13}$CO ($J$ = 1--0) spectral line observations towards the north-western region of the Mon R2 HFS obtained with the Nobeyama 45-m telescope to investigate the gas flow along and across the filaments (Fs) and inter-filamentary regions (IFs), respectively. 

$\bullet$ We identify the velocity coherent structures using multi-Gaussian fitting and the friends-of-friends algorithm from both spectral line data. We find three dense Fs (F1, F2, and F3) and three IFs (IF1, IF2, and IF3) (Figure \ref{fig:velgrad}).

$\bullet$ We estimate total mass accretion rates along and across Fs and IFs. The mass accretion rates along Fs and IFs range from \jh{17.7} to \jh{20.8} M$_\odot$ Myr$^{-1}$ and are greater than those across Fs and IFs, which range from \jh{6.6} to \jh{11.9} M$_\odot$ Myr$^{-1}$ (Table \ref{tab:para}).

$\bullet$ To compare the mass accretion rates across and along Fs and IFs independently from their total mass, we divide them into sub-regions with spacings of radial distance of 0.2 pc. Then, we compare the mean local velocity gradients and mass accretion rates along and across Fs and IFs. The mean local velocity gradients (mass accretion rate) along Fs are larger than those in the IFs suggesting that the dense gas along the filments is infalling faster towards the hub than the lower density gas in the inter-filamentary region. 

$\bullet$ The observed mass flow along the IF onto the hub points to the important gas mass reservoir, which is located in the low density gas of the inter-filament region. We found that the total mass infalling into the hub increases by \jh{1.5} times when considering the gas mass in the IF. This result is obtained by extrapolating the mass accretion rate measured within the 59$^\circ$ sector of the Mon R2 HFS to the full 360$^\circ$ region. Thus, we show that the mass accretion rates in IFs are comparable to those in Fs within our analyzed region. Therefore, we suggest that considering the total gas mass, both low- and high-density, is important to understand the role of HFS in promoting the formation of massive stars within the hub. 



$\bullet$ We also investigate the core mass functions (CMF) within and outside the hub. The slope of the CMF inside the hub is approximately eight times shallower than that outside the hub, reflecting an increase of the number and the mean mass of cores in the hub. The cores inside the hub are more massive and warmer than those outside the hub. 

Our study towards the Fs and IFs of the Mon R2 HFS shows dominant gas flows toward the hub, which likely plays an important role in increasing the mass and density within the hub promoting star formation. We also detect significant mass flows across the inter-filamentary regions towards the dense filament that could contribute to the growth of dense filaments and the formation of cores within them. In this study, we have analyzed only the north-west part of Mon R2 HFS. To generalize our finding, we will extend the analysis to other HFS targets and we plan to compare our results with numerical simulations.

\appendix

\section{Column density and temperature maps of Mon R2} \label{sec:temp}

\begin{figure*}[htb!]
\epsscale{1.0}
\plotone{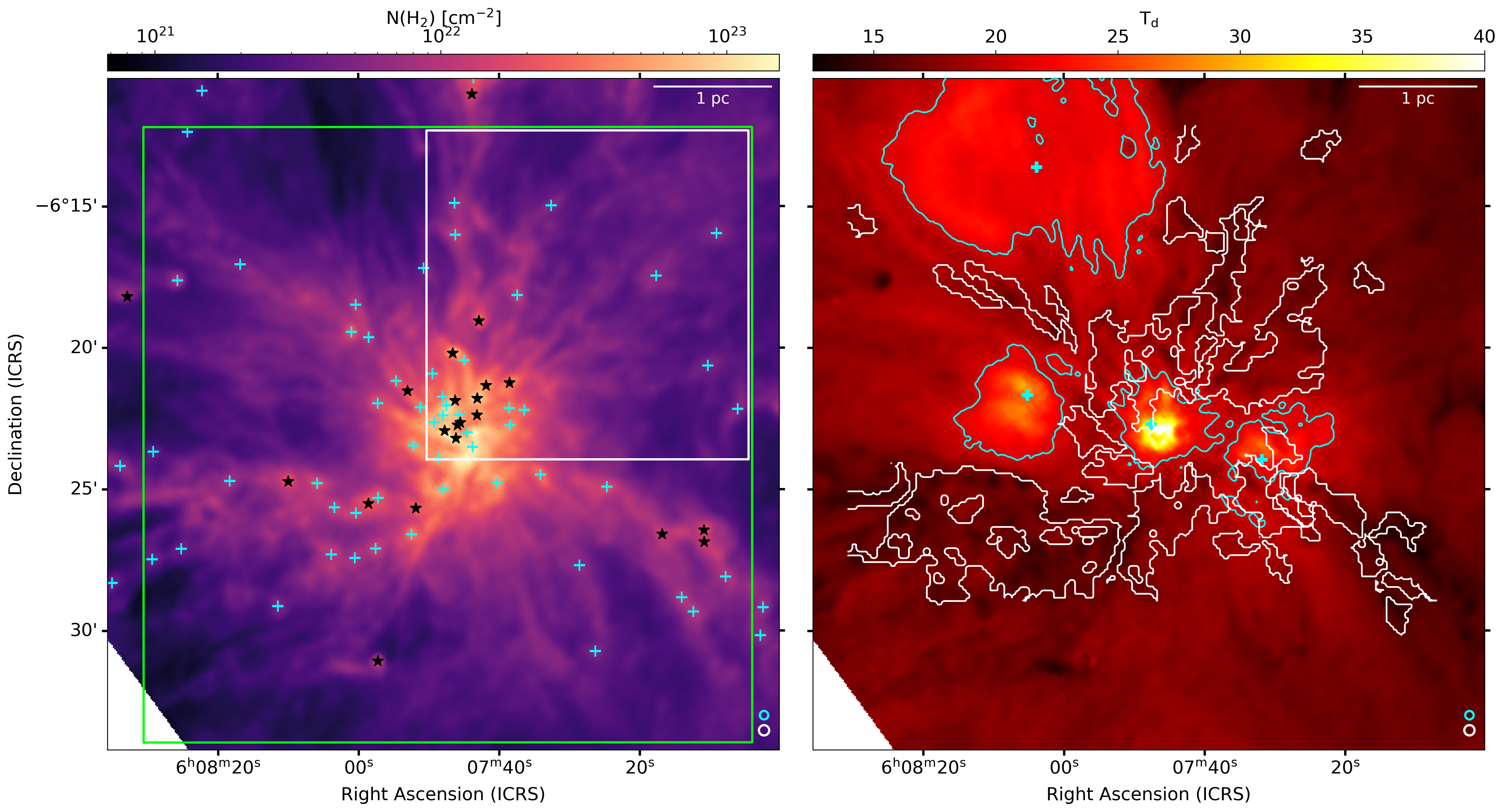}
\caption{H$_2$ column density (left) and dust temperature (right) maps obtained from $Herschel$ data \citep{Didelon2015,Kumar2021}. The green box in the left panel indicates the observed region using the Nobeyama 45 m telescope. The white box in the left panel shows the same area shown in Figure \ref{fig:velgrad}. The black stars and cyan crosses in the left panel indicate the prestellar and protostellar cores respectively defiend and analyzed in \citet{Rayner2017}. Cyan contours in the right panel correspond to a dust temperature of 21 K. Cyan crosses present the position of heating sources of {\sc{Hii}} regions \citep{Didelon2015}. White contours indicate the velocity-coherent components of C$^{18}$O emission as shown in the left panel of Figure \ref{fig:comp}. Scale bars are located at the top right corners of both panels. Cyan and white circles at the bottom right corners of both panels indicate the effective beam size of 21.8$''$ and 18$''$ for the Nobeyama 45 m telescope and the $Herschel$ observations, respectively. \label{fig:temp}} 
\end{figure*}

Figure \ref{fig:temp} shows the maps of H$_2$ column density and dust temperature obtained using $Herschel$ observations \citep{Didelon2015, Kumar2021}.
The green box on the left panel of the figure shows the region observed in the $^{13}$CO and C$^{18}$O toward Mon R2. Four {\sc{Hii}} regions are located in Mon R2 marked as cyan contours in the right panel of the figure, and their sizes and locations are described in \citet{Didelon2015}. The bright central region corresponds to the central hub of Mon R2, while the radial features trace filaments converging toward the hub. We also show that the velocity coherent structures identified in C$^{18}$O  are radial towards the center of the hub in the north-west region we analyzed (Figure \ref{fig:comp}). The region is less affected by {\sc Hii} regions. In the region, we also identify a large $^{13}$CO velcoity coherent component (Figure \ref{fig:13co}). For these reasons, we use only the northern structure for further analysis. We postpone the analysis of the entire region for a further publication. 


We estimate the H$_2$ gas mass from the $^{13}$CO column density, rather than $Herschel$ column density map. \jh{The column density of $^{13}$CO is calculated using the same equation in \citet{Trevino2019},}
\begin{equation}
N(^{13}\mathrm{CO}) = 4.69 \times 10^{13}T_\mathrm{ex} e^{\frac{5.3}{T_\mathrm{ex}}} \int T_\mathrm{mb} dv,
\end{equation}
where $T_\mathrm{ex}$ is the excitation temperature, \jh{and $\int T_\mathrm{mb} dv$ is the integrated intensity of $^{13}$CO component.} We assume that $T_\mathrm{ex}$ is equal to the dust temperature derived from $Herschel$ data \jh{ranging from 15-19 K} \citep{Didelon2015}.
\jh{The H$_2$ gas mass is given by 
\begin{equation}
M_{\mathrm{H}_2} = \frac{N(^{13}\mathrm{CO})}{X}A\mu m_\mathrm{H}
\end{equation}
where $X$ is the abundance ratio of [$^{13}$CO/H$_2$] of 1.7 $\times$ 10$^{-6}$ \citep[e.g.,][]{Ginard2012}, $A$ is the surface area, $\mu$ is the mean molecular weight of 2.8, and m$_\mathrm{H}$ is the mass of hydrogen atom.}



\section{Identification of Velocity Coherent Structures}
\label{sec:velstruc}

In Mon R2, we identify extended filamentary structures and extract velocity-coherent structures using a friends-of-friends (fof) algorithm applied to the C$^{18}$O and $^{13}$CO data. Before applying the fof algorithm, we decompose multiple velocity components in each pixel using the Python toolkit $FUNStools$\footnote{\url{https://github.com/radioshiny/funstools}}. This toolkit has been applied to molecular line data of several star-forming clouds \citep{Chung2019, Chung2021, Yoo2023, Hwang2026, Moharana2026}. In the first step, the number of Gaussian components and their centroid velocities are determined using the function of $find\_peaks$ in the $scipy$ package. Based on these initial estimates, multiple-Gaussian fitting is then performed using the $curve\_fit$ function of Python.

After decomposing the C$^{18}$O spectra into multiple components, we apply the fof algorithm to extract velocity coherent structures. The fof algorithm is defined using two criteria: centroid velocity and peak brightness temperature. We select the brightest component as the initial seed and connect neighboring components to this seed when the difference in their centroid velocities is smaller than 0.069 km s$^{-1}$ (the half of a spectral resolution of 0.11 km s$^{-1}$) and when the ratio of the two peak brightness temperatures is within 15\%. This fof algorithm was used in \citet{Hwang2026}. This procedure is iteratively repeated by updating the seed with each newly connected neighboring component until no further components satisfy the criteria. In this way, a single velocity-coherent group is constructed. We then select the brightest component among the remaining unassigned components as the next seed to identify the second velocity coherent group. Repeating this procedure allows us to organize all spectral components of C$^{18}$O into velocity-coherent groups.

\begin{figure*}[htb!]
\epsscale{1.2}
\plotone{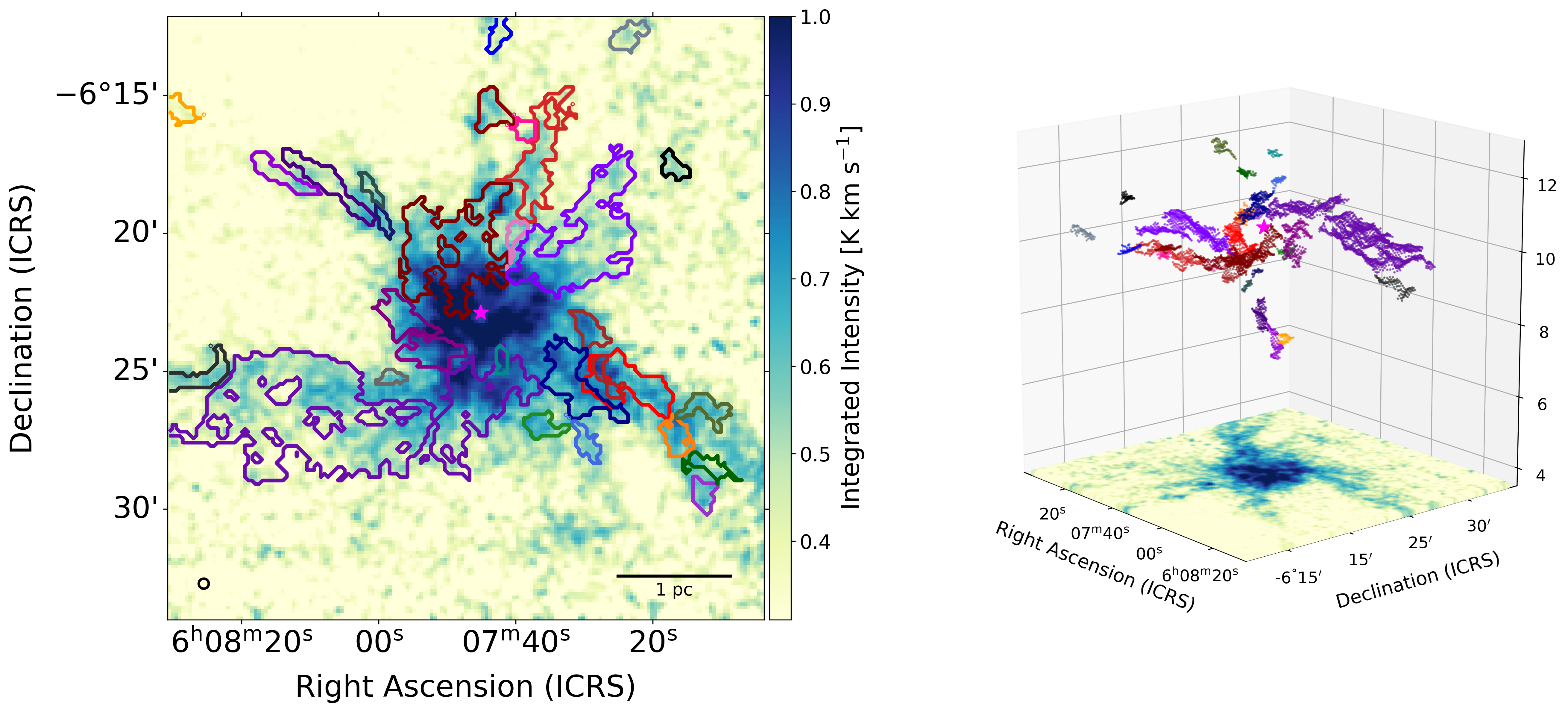}
\caption{(Left) The map of integrated intensity of C$^{18}$O integrated over velocities from 8 km s$^{-1}$ to 13 km s$^{-1}$.   Velocity-coherent components of C$^{18}$O emission shown as color contours. The magenta star indicates the position of the IRS 1 source, we consider, in our analysis, as the center of the hub of Mon R2. The effective beam size of 21.8$''$ of the Nobeyama C$^{18}$O observations is shown as black circle at the lower left corner. The physical scale bar of 1 pc at the distance of 830 pc is shown at the lower right corner. (Right) The position-position-velocity map of velocity coherent structures of which colors are the same with those in the left panel. This is available online as an interactive figure. \label{fig:comp}} 
\end{figure*}

We identify 30 C$^{18}$O velocity coherent structures with areas larger than three beam sizes (18 pixels) for further analysis. Figure \ref{fig:comp} presents these velocity coherent structures as colored contours overlaid on the integrated intensity map of C$^{18}$O. We also identify 597 C$^{18}$O velocity coherent clump-like structures with areas smaller than three beam sizes, $\sim$60\% of them located in the hub. All the identified velocity coherent structures within the hub are clump-like and not filamentary. Therefore, this work focuses on studying the dynamics of the elongated filamentary structures outside the hub, for a radial distance of 0.8 pc from the hub (Figure \ref{fig:comp}).

\begin{figure*}[htb!]
\epsscale{1.1}
\plotone{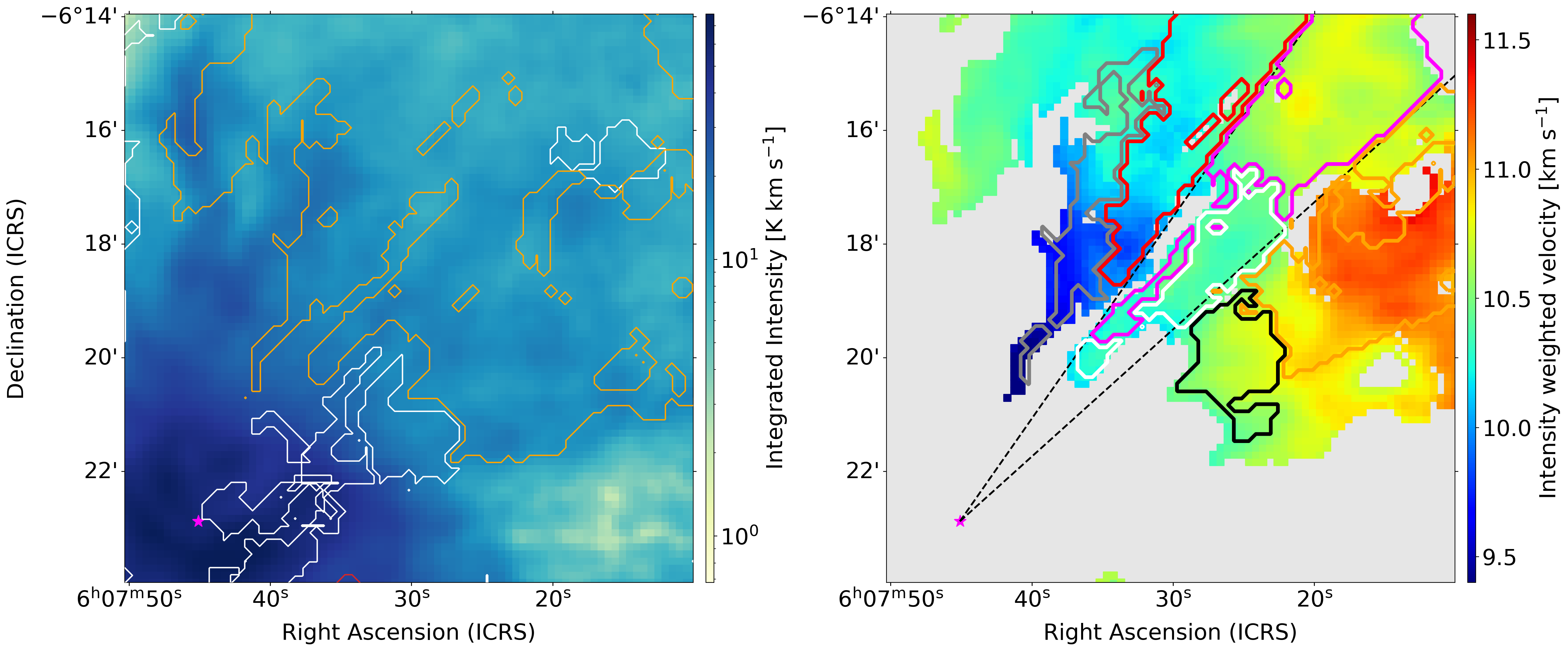}
\caption{(Left) Map of the $^{13}$CO integrated intensity integrated over velocities from 8 km s$^{-1}$ to 13 km s$^{-1}$. The white and orange contours show velocity-coherent components of $^{13}$CO emission with sizes larger than three beam sizes and the largest component, respectively. (Right) Map of intensity weigthed velocity of $^{13}$CO in the orange contours of the left panel. The colored contours represents Fs and IFs as shown in Figure \ref{fig:comp}. The two black dashed lines indicate the position angle of 42$^\circ$ and 55$^\circ$ from west to north, which correspond to the criteria to separate between IF1, IF2, and IF3 (see Section \ref{sec:fila}) \label{fig:13co}}
\end{figure*}

We also apply the fof algorithm to the $^{13}$CO data to investigate the velocity gradients in the less dense regions and to compare them with those traced by C$^{18}$O. We fit multiple Gaussian components to the $^{13}$CO spectra and then apply the fof algorithm to these components. We adopt a smaller centroid velocities threshold of 0.028 km s$^{-1}$ (instead of 0.069 km s$^{-1}$ used for C$^{18}$O data). Due to the broader linewidth of the $^{13}$CO emission, using a larger threshold for the velocity difference would connect most of the emission into a single velocity component structure. This procedure identifies 51 velocity coherent structures with areas larger than three beam sizes in Mon R2. Figure \ref{fig:13co} shows the velocity coherent structures located in the north-west region. We also find 22 structures having sizes smaller than three beams and larger than one beam in the region.


In the north-west region we analyzed in this work, our fof algorithm identified a single large structure from $^{13}$CO data (orange contours in Figure \ref{fig:13co}), while seven velocity coherent structures were identified from C$^{18}$O data in the $^{13}$CO structure (cf. Figure \ref{fig:comp}). We divided the region into Fs where C$^{18}$O and $^{13}$CO were detected and IFs, where only $^{13}$CO was detected. In addition, we divided the $^{13}$CO emission between F1 and F2 as IF1 and IF2 where the border between IF1 and IF2 is defined as a change in the velocity gradient in the azimuthal direction from west to north, as can be seen in Figure \ref{fig:13co}. Thus, we identified three IFs based on three different ranges of position angle (IF1: 55$^\circ$-67$^\circ$, IF2: 42$^\circ$-55$^\circ$ IF3: 23$^\circ$-42$^\circ$).


\section{Velocity gradient along the arc}\label{sec:arc}

\begin{figure*}[htb!]
\epsscale{1.1}
\plotone{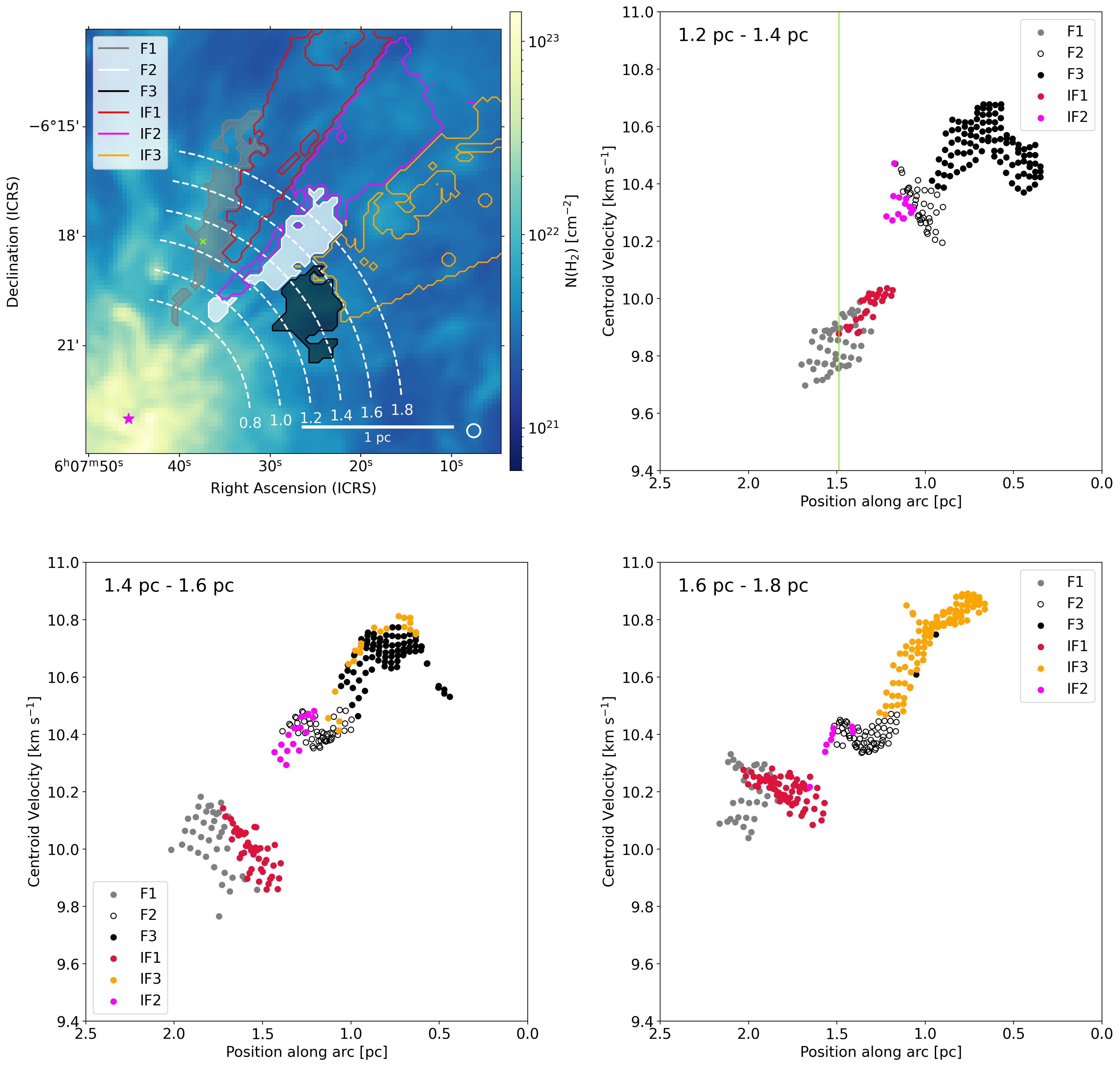}
\caption{(Top Left) Filaments and inter-filamentary regions overlaid on the H$_2$ column density map obtained from $Herschel$ data (Same as in Figure \ref{fig:velgrad}). The white dashed lines indicate the boundary of sub-regions we analyze in steps of 0.2 pc. (Top Right, Bottom Right, and Bottom left) Centroid velocities as a function of position along the arc from west (0 pc) to north (2.5 pc), within the region between two radial distances from the hub center. The two radial distances are given in the upper left corners in each panel. 
The green vertical line indicates the position of the prestellar core located in F1. \label{fig:arc}} 
\end{figure*}

In Figure \ref{fig:arc}, we show the arc lines between 0.8 pc and 1.8 pc of the radial distances. In three panels of the figure, we also show the centroid velocities as a function of the position along the arc in three ranges of radial distances, 1.2 pc - 1.4 pc, 1.4 pc - 1.6 pc, and 1.6 pc - 1.8 pc, respectively.
The position along an arc is calculated as $\theta \times r$, where $\theta$ is the angle counterclockwise from the west to north based on the declination of IRS1 as 0$^\circ$ and $r$ is the radial distance of a centroid velocity. The centroid velocities in F3 in a range of radial distance of 1.2 -- 1.4 pc shows a lambda-shape, which may trace filament formation or accretion \citep[see, e.g.,][]{Arzoumanian2018}.

The derived velocity gradient in IF1 is 0.56 $\pm$ 0.07 km s$^{-1}$ pc$^{-1}$. We estimate the mass of the velocity coherent structure of IF1 within the arc region and derive a corresponding mass accretion rate of approximately 8 M$_\odot$ Myr$^{-1}$ in IF1.

\begin{figure*}[htb!]
\epsscale{1.0}
\plotone{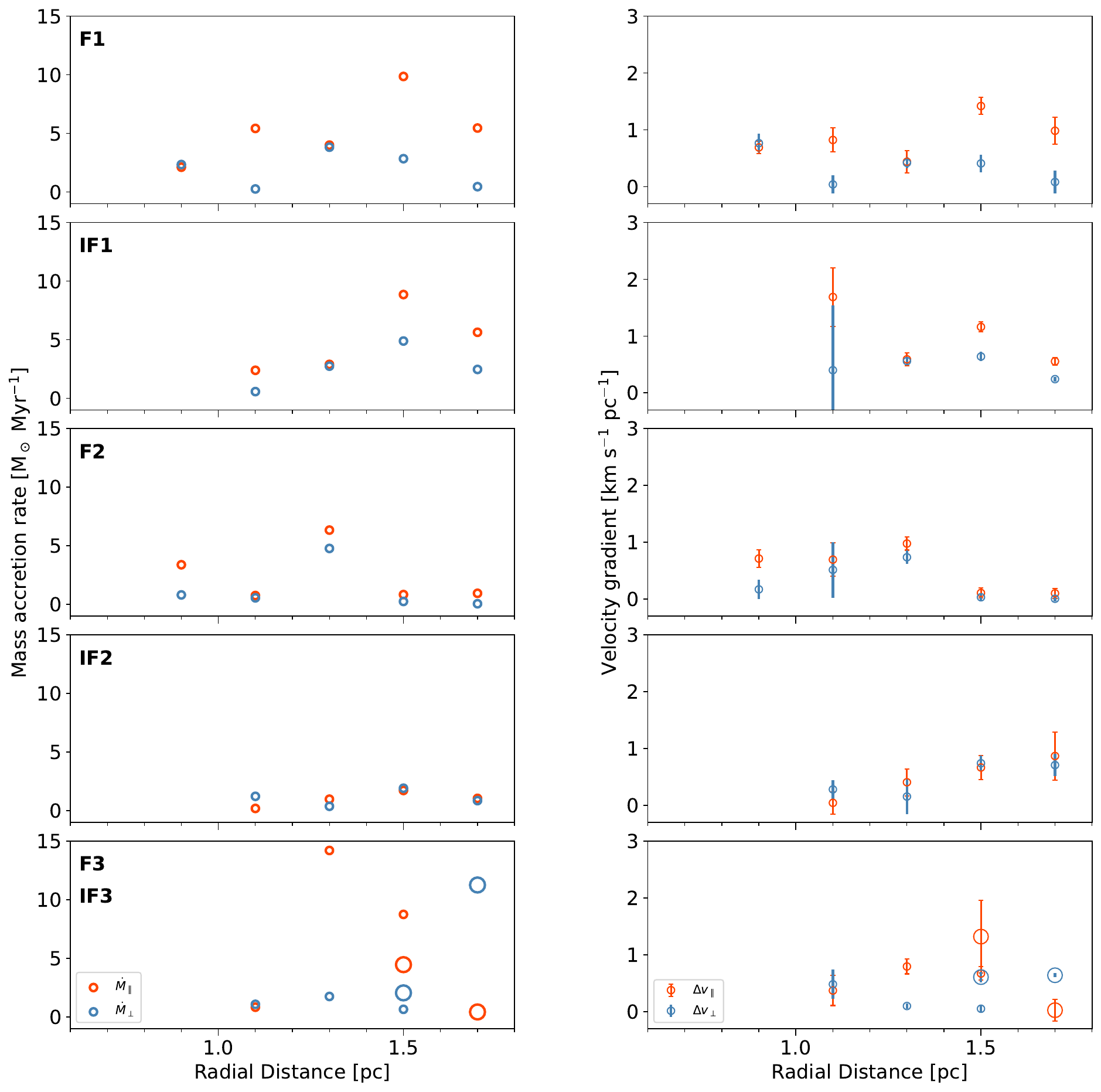}
\caption{Mass accretion rates ($\dot{M}$; Left) and velocity gradients ($\Delta v$; Right) along ($\parallel$) and across ($\perp$) the Fs and IFs measured in 0.2 pc steps (within the arcs shown in Figure \ref{fig:arc}) from the hub center. Red open circles in the left and right panels indicate $\dot M _\parallel$ and $\Delta v _\parallel$, respectively. Blue dots in the left and right panels indicate $\dot M _\perp$ and $\Delta v _\perp$, respectively. The error bars in the right panel are the fitting errors. In the last row, we overplot the values of F3 and IF3 as they are identified as a continuous structure radially extending away from the hub. The larger circles indicate values in IF3. \label{fig:arc2}} 
\end{figure*}

Figure \ref{fig:arc2} shows the mass accretion rates and velocity gradients along and across the structures  divided as the radial distance (white lines in Figure \ref{fig:arc}). When a F is compared with the nearby IF, both the mass accretion ratio and the velocity gradient are usually larger in F than IF. The mass accretion rates (velocity gradients) within most of the sub-regions in F1, IF1, and F2 are higher than across them. However, the mass accretion rates (velocity gradients) along and across IF2 are comparable. At the last point of IF3 in Figure \ref{fig:arc2}, mass accretion rates (velocity gradients) across the structures are more dominant than along them.   

\section{Mass infall rate}\label{sec:infall}

We estimate the mass infall rate ($\dot{M}$) of the gravitationally bounded prestellar core shown in Figure \ref{fig:velgrad}. The mass ($M$) and reference size ($2R$) of the prestellar core are 2.2 M$_\odot$ and 0.04 pc in \citet{Rayner2017}. The mean density of the prestellar core is given by
\begin{equation}
\rho=\frac{3M}{4\pi R^3},
\end{equation}
and the free-fall time ($t_{\rm ff}$) is calculated by 
\begin{equation}
t_{\rm ff} = \sqrt{\frac{3\pi}{32 G \rho}},
\end{equation}
where $G$ is the gravitational constant. The estimated free-fall time is 3.2 $\times$ 10$^4$ yr. The mass infall rate \jh{($M$/$t_{\rm ff}$)} is estimated as 68 M$_\odot$ Myr$^{-1}$.

\begin{acknowledgments}
The 45-m radio telescope is operated by Nobeyama Radio Observatory, a branch of National Astronomical Observatory of Japan.
This research is partially supported by the NAOJ University Support Expenses (FY2025). Data analysis was in part carried out on the Multiwavelength Data Analysis System operated by the Astronomy Data Center (ADC), National Astronomical Observatory of Japan.
This research has made use of data from the Herschel HOBYS project (http://hobys-herschel.cea.fr). HOBYS is a Herschel Key Project jointly carried out by SPIRE Specialist Astronomy Group 3 (SAG3), scientists of the LAM laboratory in Marseille, and scientists of the Herschel Science Center (HSC).
\end{acknowledgments}

\facilities{Nobeyama 45m telescope}



\end{document}